\documentclass[12pt]{article}
\usepackage{amsxtra,amssymb,amsthm,amsmath,latexsym}

\theoremstyle{plain}

\newtheorem{theorem}{Theorem}[section]
\newtheorem{lemma}{Lemma}[section]

\numberwithin{equation}{section}

\def\nd{\noindent}

\def\oH{\buildrel\circ\over H}
\def\oH1{\buildrel\circ\over H\kern-.02in{}^1}
\def\qed{{\hfill $\Box$}}

\def\const{\hbox{\,const\,}}

\begin{document}

Intern. Jour. of Diff. Eqs and Appl., 2, N3, (2001) 

%\begin{titlepage}

\title{New proof of Weyl's theorem                  
   \thanks{key words: limit circle, limit point, Weyl's solution. }
   \thanks{Math subject classification: 34B25, 34B20}
}

\author{
A.G. Ramm\\
 Mathematics Department, Kansas State University, \\
 Manhattan, KS 66506-2602, USA\\
ramm@math.ksu.edu\\
http://www.math.ksu.edu/\,$\widetilde{\ }$\,ramm}

\date{}

\maketitle\thispagestyle{empty}

\begin{abstract}
Let $lu = -u^{\prime \prime} + q(x)u$, where $q(x)$ is a real-valued
$L^2_{loc}(0, \infty)$ function. H. Weyl has proved in 1910 that for any $z$,
$Imz \neq 0$, the equation $(l - z)w=0$, $x>0$, has a solution
$w \in L^2(0, \infty)$.

We prove this classical result using a new argument.
\end{abstract}

%\end{titlepage}

\section{Introduction}
Let $lu=-u^{\prime \prime} + q(x) u$, where $q(x) \in L^2_{loc}$
is a real-valued function. Fix an arbitrary complex number $z$, $Im z >0$,
and consider the equation
\begin{equation} \label{1.1}%1.1 
        lw -zw =0, \quad x > 0
        \end{equation}

H. Weyl proved \cite{5} that equation \eqref{1.1} has a solution
$w \in L^2(0, \infty)$, which is called a Weyl's solution. He gave the
limit point-limit circle classification of
the operator $l$: if equation \eqref{1.1} has only one solution
$w \in L^2(0, \infty)$, then it is a limit point case, otherwise it is a
limit circle case.

Weyl's theory is presented in several books, e.g. in \cite{4}, \cite{3}.
This theory
is based on some limiting procedure $b \to \infty$ for the solutions to
\eqref{1.1} on a finite interval $(0, b)$. In \cite{3} a 
nice different proof
is given for continuous $q(x)$.

The aim of our paper is to give a new method 
for a proof of Weyl's result. 

\begin{theorem} %1.1
Equation \eqref{1.1} has a solution $w \in L^2(0, \infty)$.
\end{theorem}

Let us outline the new approach and the steps of the proof.

Since $q(x)$ is a real-valued function, 
symmetric operator $l_0$ defined on
a linear dense subset $C^\infty_0 (0, \infty)$
of $H=L^2 (0, \infty)$ by the expression
$lu = -u^{\prime \prime} + q(x) u$ has a selfadjoint 
extension, which we
denote by $l$. Therefore the resolvent 
$(l - z)^{-1}$ is a bounded linear
operator on the Hilbert space
$$H=L^2(0, \infty), \quad \|(l-z)^{-1} \| \leq |Imz|^{-1}.$$

This operator is an integral operator with 
the kernel $G(x,y;z)$, which is a
distribution satisfying the equation
\begin{equation}\label{1.2} %1.2
        (l-z) G(x,y;z) = \delta(x-y),\quad G(x,y;z) = G(y,x;z).
        \end{equation}

We will prove that
\begin{equation}\label{1.3} %1.3
        \int^\infty_0 \left| G(x,y;z) \right|^2 dy \leq c(x;z) \quad
        \forall x \in (0, \infty), \quad Imz >0,
        \end{equation}
where $c(x;z) = \const >0$.

The kernel $G(x,y;z)$, which is the 
Green function of the operator $l$, can
be represented as
\begin{equation}\label{1.4}%1.4
        G(x,y;z) = \varphi (y;z) w(x;z), \quad x>y,
        \end{equation}
where $w$ and $\varphi$ are 
linearly independent solution to \eqref{1.1}, so
that $w(x;z)\not\equiv 0$.
From \eqref{1.3} it follows that
\begin{equation}\label{1.5}%1.5
        w(x;z) \in L^2(0, \infty).
        \end{equation}

A detailed proof is given in section 2.

One may try to prove the existence of a Weyl's solution as follows:
take an $h\in L^2_{loc}(0,\infty)$, $h=0$ for $x>R$, $h\not\equiv 0$,
and let $W:=W(x,z):=(l-z)^{-1}h$, $Im z>0$.
 Then $W$ solves (1.1) for
$x>R$
and $W\in L^2(0,\infty)$ since $l$ is a selfadjoint operator in $H$.
However, one has to prove then that $W$ does not vanish identically 
for $x>R$, and this will be the case not for an arbitrary $h$
with the above properties. 
In our paper the role of
$h$ is played by the delta-function, and since $ \varphi (y;z)$
and $w$ in (1.4) are linearly independent solutions of (1.1), one
concludes that $w$ does not vanish identically.

\section{Proofs}
\begin{lemma} %1
If $q(x) \in L^1_{loc} (0, \infty)$ and $q(x)$ is real-valued, then symmetric
operator
$$l_0 u := -u^{\prime \prime} + q(x) u,
  \quad D(l_0) = \left\{ u : u \in C^\infty_0 (0, \infty), \quad
  l_0 u \in H := L^2 (0, \infty) \right\}$$
is defined on a linear dense in $H$ subset, and admits a selfadjoint
extension $l$.
\end{lemma}

\begin{proof}
This result is known: the density of the domain of definition of the
symmetric operator $l_0$ mentioned in Lemma 1 and the
existence of a selfadjoint extension are proved in \cite{2}. The
defect
indices of $l_0$ are (1,1) or (2,2), so that by von Neumann extension
theory $l_0$ has selfadjoint extensions (see \cite{2}). 
Actually we assume in the
Appendix that $q \in L^2_{loc} (0, \infty)$, in which case the conclusion
of Lemma 2.1 is obvious: $C^\infty_0 (0, \infty)$ is the linear dense
subset in $H$ on which $l_0$ is defined.
\end{proof}

Let $l$ be a selfadjoint extension of $l_0, (l-z)^{-1}$
be its resolvent, $Imz >0$, and $G(x,y;z)$ be the resolvent's
 kernel (in the sense of
distribution theory) of $(l-z)^{-1}$, $G(x,y;z) = G(y,x;z)$.

\begin{lemma}%2
For any fixed $x \in [0,\infty)$ one has
\begin{equation}\label{2.1} %2.1
        \left(\int^\infty_0 \left| G(x,y;z)\right|^2 dy \right)^{\frac{1}{2}}
        \leq c, \quad c = c(x;z) = \const >0.
        \end{equation}
\end{lemma}

\begin{proof}
Let $h \in C^\infty_0 (0, \infty)$ and $u:= (l-z)^{-1} h$,
so
\begin{equation} \label{2.2}
        u(x;z) = \int^\infty_0 G(x,y;z) h(y) dy, \quad (l-z) u =h.
        \end{equation}

Let us prove that:
\begin{equation}\label{2.3}
        \left| u(x;z) \right| \leq c(x;z) \|h \|,
        \end{equation}
where $x \in [0, \infty)$ is an arbitrary fixed point, $c(x) = \const >0$,
$\| h \| := \| h \|_{L^2(0, \infty)}$, $(u, v) := (u,v)_{L^2(0, \infty)}$.

If \eqref{2.3} is proved, then
\begin{equation} \label{2.4}
        \left| \left(G(x,y;z), h \right) \right| \leq c(x;z) \| h \|.
        \end{equation}

From \eqref{2.4} the desired conclusion \eqref{2.1} follows 
immediately by the Riesz theorem about linear functionals in
$H$.

To complete the proof, one has to prove estimate \eqref{2.3}.

This estimate follows from the inequality:
\begin{equation} \label{2.5}
        \| u \| _{C(D_1)} \leq c \left(\| -u^{\prime \prime} +
        q(x)u-zu \|_{L^2(D_2)} + \| u \|_{L^2(D_2)} \right) \leq c
        \left( 1 + \frac{1}{|Imz|} \right) \| h \|,
        \end{equation}
where $c = c(D_1, D_2) = \const >0$, $D_1 \subset D_2$,
$D_2 \subset [0, \infty)$, $D_1$ is a strictly inner open subinterval of
$D_2$.

Indeed, since $l$ is selfadjoint, \eqref{2.2} implies:
\begin{equation} \label{2.6}
        \| u \| \leq \frac{\| h \|}{|Imz|}.
        \end{equation}

Moreover
\begin{equation} \label{2.7}
        -u^{\prime \prime} + qu - zu =h,
        \end{equation}
so, using \eqref{2.6}, one gets:
\begin{equation} \label{2.8}
        \| u \|_{L^2(D_2)} + \| - u^{\prime \prime} + qu - zu \|_{L^2(D_2)}
        \leq \frac{\|h \|}{|Imz|} + \| h \| \leq
        \left( 1+ \frac{1}{|Imz|} \right) \|h \|,
        \end{equation}

From \eqref{2.5}, \eqref{2.6} and \eqref{2.8} one gets \eqref{2.3}.

Let us finish the proof by proving \eqref{2.5}.

In fact, inequality \eqref{2.5} is a particular case of the well-known
elliptic estimates (see e.g. [1, pp. 239-241]), but an elementary proof
of \eqref{2.5} is given below in the Appendix.

Lemma 2 is proved.
\end{proof}

\nd{\bf Proof of Theorem 1.1}

Equation \eqref{1.2} implies that
$$G(x,y;z)= \varphi(x;z) w(y;z), \quad y \geq x,$$
where $w(y;z)$ solves \eqref{1.1}, and the function $\varphi (x;z)$ is also
a solution to \eqref{1.1}. Inequality \eqref{2.1} implies
$w \in L^2(0, \infty)$ if $Imz>0$.

Theorem 1.1 is proved.
\qed

To make this paper self-contained we give an elementary proof of inequality
\eqref{2.5} in the Appendix. This proof allows one to avoid reference to the
elliptic inequalities \cite{1}, the proof of which in \cite{1} is long and
complicated (in \cite{1} the multidimensional elliptic equations of general
form are studied, which is the reason for the complicated argument in
\cite{1}).

\nd{\bf Appendix:}
An elementary proof of inequality \eqref{2.5}.

Since $u(x)$ is $C^1_{loc} (0, \infty)$ it is sufficient to prove
\eqref{2.5} assuming that $D_1 = (a,b)$ and $b-a$ is arbitrarily small.
Let $\eta (x) \in C^\infty_0 (a,b)$
be a cut-off function,
$0 \leq \eta \leq 1$, $\eta (x) = 1$ in $(a+ \delta, b - \delta)$,
$0 < \delta < \frac{b-a}{4}$,
$\eta (x) = 0$
in a neighborhoods of points $a$ and $b$.

Let $v = \eta u$. Then \eqref{2.2} implies:
$$l v = \eta h - 2\eta^\prime u^\prime - \eta^{\prime \prime} u, \quad
  v(a) = v^\prime (a) = 0 .$$
Thus
%\begin{equation}
$$ \label{A.1}
        v^{\prime \prime} = qv - zv - \eta h + \eta^{\prime \prime} u +
        2 \eta^\prime u ^\prime,
        \eqno{(A.1)}$$
        %\end{equation}
and
\begin{align}
    \label{A.2}
        \left| v(x) \right| = 
         &\left| \int^x_a (x-s) v^{\prime \prime} (s)
        ds \right| \leq c_1 \int^b_a
        \left[ |qv| + |z| |v| \right] ds + c_2,  \notag \\
         &\int^b_a |h| ds +
        c_2 \int^b_a |u| ds + c_2
        \int^b_a |u^\prime | ds.
        %\eqno{(A.2)}
        \tag{A.2}
        \end{align}
        
Here
$$c_1 = b-a, c_2 = \max_{a \leq x \leq b} \left[ |\eta (x)| +
  |\eta^{\prime \prime} + 2|\eta^\prime | \right].$$

If $b-a$ is sufficiently small, then
$$c_1 \int^b_a \left( |q| + |z| \right) dx \max_{a \leq x \leq b} |v(x)| <
  \gamma \max_{a \leq x \leq b} |v(x)|, \quad 0 < \gamma < 1.$$

Therefore (A.1) implies
%\begin{equation}
$$\label{A.3}
        \max_{a \leq x \leq b} |v(x)| \leq c_3 \left[ \| h \|_{L^2(a,b)} +
        \| u \|_{L^2(a,b)} = \| u^\prime \|_{L^2(a,b)} \right],
        \eqno{(A.3)}$$
        %\end{equation}
where $c_3 = c_3 (a,b;z)$. From (A.3) and \eqref{2.6} it follows that
inequality \eqref{2.5} holds, provided that:
%\begin{equation}
$$ \label{A.4}
        \| u^\prime \|_{L^2(a,b)} \leq c \| h \| + \delta \| u \|_{L^\infty}.
        \eqno{(A.4)}$$
        %\end{equation}

The last estimate is proved as follows. Multiply \eqref{2.2} by
$\eta \overline u$ (the bar stands for complex conjugate and $\eta$ is
a cut-off function, $\eta \in C^\infty_0 (a,b)$ and integrate over
$(a,b)$ to get
$$\int^b_a |u^\prime|^2 \eta dx = \int^b_a u^\prime \overline u \eta^\prime dx
  + \int^b_a \eta h \overline u dx + z \int^b_a \eta |u|^2 dx - \int^b_a
  q|u|^2 \eta dx := I_1 + I_2 + I_3 + I_4.$$

One has, using the inequality
$|uv| \leq \varepsilon |u|^2 + \frac{|v|^2}{4 \varepsilon} , \varepsilon >0$,
$$|I_1| \leq c \left( \varepsilon \|u^\prime \|^2 +
  \frac{1}{4 \varepsilon} \| u\|^2 \right), \quad
  c = \max |\eta^\prime| ,$$
$$|I_2| + |I_3| \leq c \left( \| h \| \| u \| + \| u \|^2 \right) \leq
  c_1 \| h \|^2,$$
where \eqref{2.6} was used,
$$|I_4| \leq \| qu\| \| u \| \leq \|q \|_{L^2} \| u \|_{L^\infty}
  \| u \|. $$

Thus, if $a < a_1 < b_1 < b$, where $\eta = 1$ on $[a_1, b_1]$, one gets
%\begin{equation}
$$\label{A.5}
        \int^{b_1}_{a_1} | u^\prime|^2 dx \leq C \left( \| h \|^2 +
        \| u \|_{L^\infty} \| h \| \right) \leq \delta \| u \|^2_{L^\infty}
        + C \| h \|^2,
        \eqno{(A.5)}$$
        %\end{equation}
where $C=C( \varepsilon, z, a, b, \delta) = \const >0$, $0 < \delta$
can be chosen arbitrarily small. Inequality (A.5) implies
(A.4).

Inequality \eqref{2.5} is proved. \qed

\end{document}